\documentclass[pra,aps,twocolumn,showpacs]{revtex4}
\usepackage{bm}
\begin{document}

\title{Unitary Gate Synthesis for Continuous Variable Systems}

\author{Jarom\'{\i}r Fiur\'{a}\v{s}ek}
\affiliation{Ecole Polytechnique, CP 165, 
Universit\'{e} Libre de Bruxelles, 1050 Brussels, Belgium }
\affiliation{Department of Optics, Palack\'{y} University, 
17. listopadu 50, 77200 Olomouc, Czech Republic}

\begin{abstract}
We investigate the synthesis of continuous-variable two-mode unitary 
gates in the setting where two modes $A$ and $B$ are coupled 
by a fixed quadratic Hamiltonian $H$. The gate synthesis
consists of a sequence of evolutions governed by Hamiltonian $H$ 
interspaced by local phase shifts applied to $A$ and $B$. We concentrate on
protocols that require the minimum necessary number of steps and we show how 
to implement the beam splitter and the two-mode squeezer in just three steps. 
Particular attention is paid to the Hamiltonian $x_A p_B$ that describes the
effective off-resonant interaction of light with the collective atomic spin.
\end{abstract}


\maketitle

\section{Introduction}

One of the central problems  of the quantum information theory is to establish
what resources are sufficient for universal quantum computation.
In this context, the question whether a given Hamiltonian $H$ can 
{\em simulate } another one has attracted  considerable attention recently 
\cite{Dur01,Bennett02,Dodd02,Nielsen02,Wocjan02,Vidal02}.
In its simplest form, this problem may be formulated as follows. 
Consider two parties, traditionally referred to as Alice and Bob, 
possessing a single  qubit each. The interaction between those two qubits  
is governed by a fixed  Hamiltonian $H$, that is determined by the physical
properties of the systems that represent the qubits.  In addition to the
interaction $H$, Alice and Bob may attach (local) ancillas to their qubits and 
perform arbitrary local unitary operations on their subsystems. It is usually
assumed that these local operations are very fast compared to the evolution
induced by the Hamiltonian $H$. The task for Alice and Bob is to simulate the
evolution due to different Hamiltonian $H^\prime$. Two kinds of simulations should be
distinguished. The infinitesimal time simulation 
\cite{Dur01,Bennett02,Dodd02,Nielsen02,Wocjan02,Vidal02} consists of
simulating the action of the Hamiltonian $H^\prime$ for an infinitesimally 
short time interval $\Delta t$. The {\em gate synthesis}
\cite{Khaneja01,Vidal02b,Hammerer02,Masanes02,Bremner02,Zhang02} requires 
the implementation of the unitary transformation $U^\prime=\exp(-iH^\prime t)$ 
for some {\em finite} time $t$.  

It turns out that in the two-party setting all nonlocal Hamiltonians $H$ are
qualitatively equivalent. Given enough time $\tau$, Alice and Bob can, with the
help of local ancillas,  simulate the evolution $\exp(-iH' t)$ for any $H'$ 
\cite{Bennett02}. 
The central question, then, is what is the optimal simulation. The latter 
may be defined as a simulation that requires the shortest time.  
For the two-qubit 
case,  this problem has been completely solved and the optimal protocols 
for Hamiltonian \cite{Bennett02,Vidal02} and unitary gate \cite{Vidal02b} 
simulations have been determined.
The situation becomes much more complicated for higher dimensional systems
and for higher number of involved parties. Simulation 
protocols suggested for these generic settings are rather 
involved and it is not known which protocols are optimal.

It should be stressed that most of the work focused on discrete variable systems:
qubits or, more generally, qudits. Recently, however, Kraus {\em et al.} 
extended the notion of Hamiltonian simulation to continuous variable systems
\cite{Kraus02}. 
They assumed that Alice and Bob possess a single-mode system each and these 
two modes are coupled via quadratic Hamiltonian (we assume $\hbar=1$ 
throughout this paper):
\begin{equation}
H=c_{11}x_A x_B+c_{12}x_A p_B + c_{21}p_A x_B + c_{22} p_A p_B,
\label{Hdef}
\end{equation}
where $x_j$ and $p_j$ are two conjugate quadratures of the $j$th mode.
Kraus {\em et al.} showed that almost every Hamiltonian (\ref{Hdef}) 
is capable to  simulate any other  Hamiltonian of the form (\ref{Hdef}) 
provided that Alice and Bob can apply fast local phase shifts  
described by single-mode Hamiltonians $H_{A}=x_A^2+p_A^2$ 
and $H_{B}=x_B^2+p_B^2$. 

These results are interesting both  from the theoretical and experimental 
points of view. In particular,  the off-resonant 
interaction of light with the collective atomic spin
\cite{Kuzmich98,Kuzmich00,Kuzmich00theory,Duan00,Julsgaard01,Schori02,Kuzmichbook}
can be described by the  effective unitary transformation  
\begin{equation}
U=\exp(-itH_{AL}),
\label{UAL}
\end{equation}
where the Hamiltonian 
\begin{equation}
H_{AL}=\kappa x_A p_B
\label{Hxp}
\end{equation}
is a special instance of (\ref{Hdef}). The typical geometry of the experiments 
is such
that a light with strong coherent field polarized along the $x$ axis propagates
along the $z$ axis through the atomic sample, whose spin is also polarized along
the $x$ axis. The $x$ and $p$ quadratures are defined as the properly normalized  
$y$- and $z$-components of the collective spin operators describing the polarization 
state of light and atomic ensemble, respectively 
\cite{Kuzmich98,Kuzmich00theory,Duan00}. 
In recent beautiful 
experiments it was demonstrated that the interaction (\ref{UAL}) can be employed to 
squeeze  the atomic spin \cite{Kuzmich00}, 
entangle two distant atomic ensembles \cite{Julsgaard01}, and transfer 
the quantum state of light into the atoms \cite{Schori02}. Schemes for teleportation and 
swapping of the  quantum state of collective atomic spin  have been 
suggested \cite{Kuzmich00theory,Duan00}. 
These experiments and proposals in fact rely on the quantum non-demolition
(QND) measurement of the atomic quadrature, possibly accompanied by a suitable 
feedback. 

As showed by Kraus {\em et al.}, the 
Hamiltonian (\ref{Hxp}) can simulate any Hamiltonian 
(\ref{Hdef}). In particular, $H_{AL}$ can be used to implement 
a beam splitter and a two-mode squeezer. This is very appealing since
it suggests that, for instance, the  storage of the quantum state of light
into atoms and the subsequent readout of the quantum memory -- the transfer of
quantum state of atoms onto light -- can be implemented in a unitary way if the
Hamiltonian (\ref{Hxp}) is used to simulate a beam splitter. 

However, currently there are technical difficulties that will complicate 
the actual practical realization of this procedure. The effective unitary
transformation (\ref{UAL})  describes the modification of the
polarization state of the light pulse after the passage through the atomic
sample  due to its coupling with atoms. 
This means that the  Hamiltonian simulation requires 
several passages of the light pulse through the atomic sample
(c.f. the detailed description of the simulation protocol in Sec. II).
In currently envisaged experiments, the  pulse width must be at least  $1$ $\mu {\rm s}$ 
\cite{Polzikprivate}
which corresponds to the length $300$ m. This long pulse would have to 
be stored somewhere (e.g. in an optical fiber) until its tail leaves the 
atomic sample. Only then can the pulse (with properly applied phase shifts) 
be fed to the atomic sample again. 

These practical considerations imply that the approach relying on 
the infinitesimal time simulation is not very convenient from 
the experimental  point of view. It is 
possible to simulate a gate by concatenating a large sequence of 
short-time Hamiltonian simulations but this would 
require a large number of manipulations and passages of the light pulse 
through the sample. Since, in practice, every round of the gate 
synthesis procedure  is necessarily accompanied by some losses and 
other errors, the accumulation of the errors  would negatively influence 
the simulation. 

In this paper, we show how to implement several important
two-mode interactions with the Hamiltonian (\ref{Hdef}) such that the  number of 
the applications of the Hamiltonian (\ref{Hdef}) is minimized. 
We demonstrate that only three sequences 
of evolution governed by Hamiltonian $H$,  interspaced by (fast) local phase 
shifts on both subsystems, suffice to implement a two-mode squeezer and a 
beam splitter. For the specific
Hamiltonian (\ref{Hxp}) we also design a single-mode squeezing gate that 
involves four evolution steps. These results illustrate that several important 
quantum information processing tasks such as entangling the light and collective
atomic spin, or a transfer of the quantum state of light into atomic clouds
and vice versa, can be carried out with a small number of repeated passages of
the light pulse through the atomic sample.

This paper is structured as follows. In Sec. II we introduce the notation,
the canonical form of the interaction Hamiltonian (\ref{Hdef}) and we
describe the gate simulation protocol. 
 In Sec. III, we consider the simple interaction Hamiltonian 
(\ref{Hxp}) and we show how to implement the two-mode squeezing operation,
beam splitter transformation and also single-mode squeezing as a sequence of 
three (or four) intervals of evolution governed by Hamiltonian (\ref{Hxp})
combined with local phase shift operations. In  Sec. IV we extend this
analysis to the generic interaction Hamiltonians (\ref{Hdef}). 
Finally, Sec. V contains conclusions.

\section{Description of the simulation protocol}

In this section we describe the simulation protocol. 
 The gate synthesis consists of a sequence  of
$N$ intervals of evolution governed by the Hamiltonian $H$ followed by 
local unitary phase shift transformations. The resulting unitary gate $G$
is given by
\begin{equation}
G=V_N^\dagger e^{-iHt_N} V_N  \ldots V_{2}^\dagger 
e^{-iH t_{2}} V_{2}  V_{1}^\dagger e^{-iH t_1} V_{1}.
\label{Gdef}
\end{equation}
The  local phase shift operation applied to modes A and B reads
\begin{equation}
V_j=e^{-i\phi_{jA}a^\dagger a}\otimes  e^{-i\phi_{jB} b^\dagger b},
\label{Vdef}
\end{equation}
where $a$ and $b$ are the annihilation operators of modes $A$ and $B$,
respectively.
Note that $V_{j+1}^\dagger V_{j}$ is still of the form (\ref{Vdef}), with
$\phi_A=\phi_{jA}-\phi_{j+1,A}$ and $\phi_B=\phi_{jB}-\phi_{j+1,B}$.
With the help of the useful identity
\begin{equation}
U^\dagger \exp(-iHt) U= \exp( -i U^\dagger H U t)
\label{Identity}
\end{equation}
we can rewrite Eq. (\ref{Gdef}) as
\begin{equation}
G= e^{-i H_N t_N }\ldots e^{-i H_{2}t_2} e^{-iH_1 t_1},
\label{Grewritten}
\end{equation}
where $H_j=V_j^\dagger H V_j$.

The Hamiltonian (\ref{Hdef}) is characterized by four parameters. 
However, by means of local rotations, we can always transform this Hamiltonian 
to a diagonal form 
\begin{equation}
H_c= c_1 x_A p_B+c_2 p_A x_B,
\label{Hcanonical}
\end{equation}
where $c_1=\sigma_1$ and $c_2=\sigma_2{\det}[C]/|\det [C]|$, and $\sigma_1$ and
$\sigma_2$ are the singular values of the matrix $C$ defined 
as $(C)_{ij}=c_{ij}$  \cite{Kraus02}. In close analogy 
to the qubit case \cite{Bennett02}, we may refer to $H_c$
 as the  {\em canonical form}  of $H$. 
Mathematically, we have
\begin{equation}
\exp(-iH_c t)= \tilde{V}^\dagger \exp(-iHt) \tilde{V},
\label{Hcanonicaltransform}
\end{equation}  
where  $\tilde{V}$ is a local rotation (\ref{Vdef}).
This shows that, without loss of generality, we may assume that $H$ has the
canonical form (\ref{Hcanonical}). In particular, it follows that $H$ 
is able to simulate an arbitrary $H'$ (\ref{Hdef})
if and only if $H_c$ is able to simulate an arbitrary canonical Hamiltonian 
(\ref{Hcanonical}).

In Eq. (\ref{Vdef}) the phase shifts $\phi_j$ may be arbitrary. 
In what follows, we focus on the phase shifts that preserve the canonical 
form of $H$. There are four inequivalent possibilities: \\
(a) $\phi_A=0$, $\phi_B=0$,
\begin{equation}
H_1=c_1 x_A p_B+c_{2}p_A x_B.
\end{equation}
(b) $\phi_A=\pi/2$, $\phi_B=3\pi/2$,
\begin{equation}
H_2=c_2 x_A p_B+c_{1}p_A x_B.
\end{equation}
(c) $\phi_A=\pi$, $\phi_B=0$,
\begin{equation}
H_3=-c_1 x_A p_B-c_{2}p_A x_B.
\end{equation}
(d) $\phi_A=\pi/2$, $\phi_B=\pi/2$,
\begin{equation}
H_4=-c_2 x_A p_B-c_{1}p_A x_B.
\end{equation}
From the structure of these Hamiltonians we can deduce 
that  two different noncommuting
canonical Hamiltonians  $H_1$ and $H_2$ are available. Furthermore,
we can see that $H_3=-H_1$ and $H_4=-H_2$, hence we can implement any
transformations of the form $\exp(-iH_1t)$ and $\exp(-iH_2t)$ where $t$ is 
an {\em arbitrary} real number, positive or negative.
The two specific cases $c_1=c_2$ and $c_1=-c_2$ when $H_1=\pm H_2$ and the
simulation is not possible correspond to the Hamiltonians of a two-mode squeezer
and a beam splitter, respectively.

\section{XP coupling}

Having established the notation and described the simulation protocol, we may
proceed to the unitary gate synthesis. Namely, we would like to
decompose the unitary transformation $G$ that we want to simulate into a
sequence of unitary evolutions governed by Hamiltonians $H_1$and $H_2$ 
that were defined in the previous section,
\begin{equation}
G= e^{-iH_2 t_N} e^{-i H_1 t_{N-1}}\ldots e^{-iH_2 t_2} e^{-iH_1 t_1}.
\label{Gfactorization}
\end{equation}
We are particularly interested in the simulations that involve the lowest 
possible number of steps $N$, because such simulations require low number of
local manipulations in the eventual experimental implementation.

We note here that  Eq. (\ref{Gfactorization}) is an example of a decomposition 
of a group element into a product of $N$ other  group elements. 
In the present case,  the underlying  group is the symplectic group 
$\textrm{Sp}(4,R)$ of all linear canonical
transformations of the quadratures of the two modes $A$ and $B$
\cite{Simon94,Arvind95}. 
It is worth mentioning here that the related problem of a 
decomposition of the symplectic transformation  into a sequence of 
simple evolutions associated with the common passive and active linear optical
elements has been studied recently. Braunstein has shown that any $N$ mode
symplectic transformation can be implemented as a sequence  of an $N$-mode 
passive linear interferometer followed by $N$ single-mode squeezers 
and another passive interferometer --- the so-called Bloch-Messiah 
decomposition \cite{Braunstein99}. The decompositions
of this kind have also been applied to investigate the properties of  nonlinear optical
couplers \cite{Fiurasek00,Rehacek02}.

In this section, we shall consider the simplest and also the experimentally
relevant coupling between the two 
systems described by the interaction Hamiltonian (\ref{Hxp}).
Without loss of generality, we may assume that the coupling constant is equal
to unity, hence the two relevant Hamiltonians read
\begin{equation}
H_1=x_A p_B, \qquad H_2= p_A x_B.
\label{HPolzik}
\end{equation}
All the canonical Hamiltonians (\ref{Hcanonical}) have the important property 
that the $x$ and $p$ quadratures are not mutually coupled when 
we write down the Heisenberg equations of motion for $x_j$ and $p_j$. 
This means that
the  evolution of the operators 
$\bm{x}=(x_A,x_B)^T$ and $\bm{p}=(p_A,p_B)^T$ is governed by the following
linear canonical transformations:
\begin{equation}
\bm{x}_{\rm out}=\bm{S} \bm{x}_{\rm in}, \qquad 
\bm{p}_{\rm out}=\bm{R} \bm{p}_{\rm in}.
\label{xpevolution}
\end{equation}
This decoupling of $x$ and $p$ quadratures greatly simplifies the analysis.
The transformation  (\ref{xpevolution}) must preserve the canonical 
commutation relations $[x_j,p_k]=i\delta_{jk}$. From these conditions 
we can express the matrix  $\bm{R}$ in
terms of $\bm{S}$,
\begin{equation}
\bm{R}=(\bm{S}^{-1})^T,
\end{equation}
hence the evolution of $p$ quadratures is uniquely determined by the evolution
of the $x$ quadratures. 

Our task is to implement two-mode unitary gates (symplectic transformations)
as a sequence of a small number of unitary transformations 
generated  by the Hamiltonians (\ref{HPolzik}). 
The matrices $\bm{S}_1$ and $\bm{S}_2$ associated with the
unitary evolutions $U_1=\exp(-iH_1 t)$ and $U_2=\exp(-iH_2t)$ read
\begin{equation}
\bm{S}_1(t)=\left(
\begin{array}{cc}
1 & 0\\ 
t & 1
\end{array}
\right),
\qquad
\bm{S}_2(t)=\left(
\begin{array}{cc}
1 & t \\ 
0 & 1
\end{array}
\right).
\label{S12}
\end{equation}

The factorization (\ref{Gfactorization}) can be rewritten in terms 
of the  matrices $S_j$ as follows,
\begin{equation}
\bm{S}=\bm{S}_2(t_N) \bm{S}_1(t_{N-1})\ldots \bm{S}_2(t_2) \bm{S}_1(t_1),
\end{equation}
where $\bm{S}$ is the matrix associated with the gate $G$.
Since $\det \bm{S}_1=\det \bm{S}_2=1$, we are restricted 
to a three parametric subgroup of transformations $\bm{S}$ 
such that $\det \bm{S}=1$.
In what follows, we will discuss the implementation of three important gates: 
a beam splitter, a two-mode squeezer and a single-mode squeezer.

\subsection{Beam splitter}

The beam splitter operation is described by the matrix
\begin{equation}
\bm{S}_{BS}(\phi)=
\left(
\begin{array}{cc}
\cos\phi & \sin\phi \\ 
-\sin\phi & \cos\phi
\end{array}
\right).
\label{SBS}
\end{equation}
We show that this transformation can be implemented as a sequence of three
evolutions (\ref{S12}),
\begin{equation}
\bm{S}_{BS}(\phi)=\bm{S}_1(\gamma) \bm{S}_2(\beta) \bm{S}_1(\alpha).
\label{Sthree}
\end{equation}
The explicit multiplication yields
\begin{equation}
\bm{S}_{BS}(\phi)=\left(
\begin{array}{cc}
1+\alpha\beta & \beta \\ 
\alpha +\gamma(1+\alpha\beta) & 1+\gamma\beta
\end{array}
\right).
\label{SBSmultiplied}
\end{equation}
If we compare the elements of the matrices on left- and right-hand sides of Eq.
(\ref{SBSmultiplied}), we obtain a set of equations for the 
parameters $\alpha$, $\beta$ and $\gamma$ whose solution yields:
\begin{equation}
\alpha=-\tan\frac{\phi}{2}, \qquad \beta=\sin\phi, \qquad \gamma=\alpha.
\end{equation}
The parameters $\alpha$ and $\beta$  can be negative but this is not 
an obstacle as explained in the previous section since 
we can change the sign of the Hamiltonian $H_1$ or $H_2$
by $\pi$ rotation of one of the systems.
Two cases of particular importance are (i) the balanced beam
splitter ($\phi=\pi/4$), that requires $\alpha=1-\sqrt{2}$ and
$\beta=\sqrt{2}/2$, and (ii) the swap ($\phi=\pi/2$) that
exchanges the quantum states of the two systems, $\alpha=-1$, $\beta=1$.

The swap gate is closely related to the two-step protocol for mapping the state
of collective atomic spin on light that was suggested by Kuzmich and Polzik
\cite{Kuzmichbook}. In fact, their two-step protocol can be obtained 
by simply removing the last step of the present three-step swap gate. 
The removal of the third step means that 
the mapping adds some noise and the procedure is thus only approximate.
A possible  way how to improve its performance 
is to use squeezed light. For details, see \cite{Kuzmichbook}.

\subsection{Two-mode squeezer}

Let us now turn our attention to the two-mode squeezer, described by the
following matrix,
\begin{equation}
\bm{S}_{\rm TMS}(r)=\left(
\begin{array}{cc}
\cosh r & \sinh r \\ 
\sinh r & \cosh r
\end{array}
\right).
\label{STMS}
\end{equation}
Similarly as in the case of the beam splitter, we attempt to implement this
transformation as a sequence of three evolutions, c.f. Eq. (\ref{Sthree}). 
By comparison of the right-hand side of Eq. (\ref{SBSmultiplied}) with the 
matrix (\ref{STMS}) we again obtain a system  of nonlinear 
equations  for the parameters $\alpha$, $\beta$, and $\gamma$ 
having the solution
\begin{equation}
\alpha=\tanh\frac{r}{2}, \qquad \beta=\sinh r, \qquad \gamma=\alpha.
\end{equation}
Note the similarity with the results for the beam splitter, the
difference is essentially that the goniometric functions have been replaced 
by their hyperbolic counterparts. The parameters are finite for any finite $r$.
However, $\beta$  grows exponentially with $r$ and for large $r$  we have
$\beta\propto e^r$. On the other hand, for small $r$ we get $\beta \approx r$.
This implies that we may reduce the synthesis time if we implement
the two-mode
squeezing transformation as a sequence of $n$ two-mode squeezers with $r'=r/n$.
The reduction of the time is achieved at the expense of a higher number of 
steps of the gate synthesis protocol. For modest values of squeezing 
$r$, the sequence of three evolutions is advantageous since it involves the
minimum necessary number of manipulations of the systems $A$ and $B$.

\subsection{Single-mode squeezer}

After dealing with two-mode gates, let us now focus on the
single-mode gate, namely the single-mode squeezer. Since $\det \bm{S}=1$, 
the squeezing of quadrature $x_A$ will necessarily  by accompanied by
anti-squeezing of $x_B$, and vice versa,
\begin{equation}
\bm{S}_{\rm SMS}(r)=\left(
\begin{array}{cc}
e^r & 0 \\ 
0 & e^{-r}
\end{array}
\right).
\end{equation}
It turns out that in this case a sequence of three transformations (\ref{S12})
is insufficient and we must consider a sequence of four basic evolutions:
\begin{equation}
\bm{S}_{\rm SMS}(r)=\bm{S}_2(\delta)\bm{S}_1(\gamma) \bm{S}_2(\beta) 
\bm{S}_1(\alpha).
\label{SSMS}
\end{equation}
We proceed as before and derive equations for the four parameters appearing in
(\ref{SSMS}):
\begin{eqnarray*}
e^r &=& 1+\alpha\beta+\delta(\alpha+\gamma+\alpha\beta\gamma), \\
e^{-r} &=& 1+\gamma\beta, \\
0 &=& \beta+\delta(1+\gamma\beta), \\
0 &=& \alpha+\gamma(1 +\alpha\beta).
\end{eqnarray*}
This system of equations has a one-parametric class of solutions, given by
\begin{equation}
\beta=\frac{e^r-1}{\alpha}, \qquad
\gamma=-\alpha e^{-r}, \qquad
\delta=\frac{e^r(1-e^r)}{\alpha},
\end{equation}
and $\alpha$ is arbitrary but nonzero. We may choose the optimal value of
$\alpha$ that minimizes the total time $T$ needed for the implementation of the
squeezing operation,
\begin{equation}
T=|\alpha|+|\beta|+|\gamma|+|\delta|.
\end{equation}
Assuming that $r>0$ we obtain by solving $dT/d\alpha=0$ the optimal value,
\begin{equation}
\alpha=\sqrt{\frac{e^{2r}-1}{1+e^{-r}}}.
\end{equation}
In the limit of small $r$, all four parameters $\alpha$, $\beta$,
$\gamma$, and $\delta$ are proportional to $\sqrt{r}$. This stems from the
fact that the single-mode squeezing Hamiltonian $H_{SMS}=x_A p_A$ cannot be obtained as
a linear combination of the two-mode Hamiltonians $H_1$ and $H_2$ and 
only the terms of the order of $O(t^2)$ or higher in (\ref{Gfactorization}) 
may give  rise to the contribution proportional to $H_{SMS}$.

\section{Generic quadratic coupling}
In this section we shall assume that the interaction Hamiltonian has the generic
canonical form. Although the mathematical analysis will be more involved we
shall still be able to derive analytical formulas for the interaction times
characterizing the gate synthesis. Without loss of generality, we may assume
that $c_1=1$ in (\ref{Hcanonical}). We have to distinguish two 
classes of Hamiltonians giving rise to qualitatively different evolutions of 
the quadratures in the Heisenberg picture. For $c_2>0$ the dynamics resembles 
an amplifier while for $c_2<0$ we obtain oscillatory dynamics reminiscent
that of a beam splitter. We shall discuss these two cases separately. 

\subsection{Amplifier-like Hamiltonians}

Suppose first that $c_2>0$ and introduce 
a more convenient notation $c_2=s^2$, $s>0$, hence
\begin{equation}
H_1=x_A p_B + s^2 p_A x_B, \qquad
H_2=s^2x_A p_B + p_A x_B.
\label{Hplus}
\end{equation}
It is an easy exercise to derive matrices $\bm{S}_1$ and $\bm{S}_2$ 
corresponding to the unitary evolutions governed by $H_1$ and $H_2$, 
respectively,
\begin{equation}
\bm{S}_{1}^{+}(t)=
\left(
\begin{array}{cc}
\cosh(st) & s\sinh(st) \\ 
\frac{1}{s}\sinh(st) & \cosh(st)
\end{array}
\right),
\end{equation}
\begin{equation}
\bm{S}_{2}^{+}(t)=
\left(
\begin{array}{cc}
\cosh(st) & \frac{1}{s}\sinh(st) \\ 
s\sinh(st) & \cosh(st)
\end{array}
\right).
\end{equation}

In what follows we will focus on the implementation of the 
beam splitter and two-mode squeezer transformations. 
We have seen in the previous section that  these transformations could be
implemented as a sequence of three basic evolutions governed by Hamiltonians
$H_1$ or $H_2$ that were of the form (\ref{HPolzik}). 
Moreover, there was an inherent symmetry in this gate synthesis; we
have found that $\gamma=\alpha$. It turns out that these basic symmetry 
properties remain valid also for the generic Hamiltonians (\ref{Hplus}), 
and we can thus decompose the two-mode squeezing  transformation as
\begin{equation}
{\bm S}_{TMS}(r)={\bm S}_{1}^{+}(\alpha/s) \bm{S}_{2}^{+}(\beta/s)
{\bm S}_{1}^{+}(\alpha/s).
\label{threesteps}
\end{equation}
Here the parameters $\alpha=st_1$ and $\beta=st_2$ are the rescaled interaction
times. The nonlinear equations for the parameters $\alpha$ and $\beta$ are much more
complicated than before. Nevertheless, analytical results can be obtained. 
From the condition $S_{12}=S_{21}$ we get
\begin{equation}
\tanh \beta=\frac{2\sinh\alpha\cosh\alpha}%
{\cosh^2\alpha-(s^{-2}+1+s^2)\sinh^2\alpha}.
\label{betaTMS}
\end{equation}
The condition $S_{11}=S_{22}$ is satisfied due to the symmetry
($\gamma=\alpha$), and the parameter $\alpha$  can be determined from the last
independent equation $S_{12}/S_{11}=\tanh r$, which yields
\begin{eqnarray*}
\frac{2y(s+s^{-1})}
{1+y^2(s^{-2}+1+s^2)}=\tanh r ,
\end{eqnarray*}
where $y=\tanh\alpha$. This is a quadratic equation for $y$ 
whose solution reads
\begin{equation}
\tanh\alpha=\frac{s+s^{-1}-\sqrt{1+(s^{-2}+1+s^2)(\cosh r)^{-2}}}{(s^{-2}+1+s^2)\tanh r}.
\end{equation}
We have selected the root that yields the correct limit $\tanh\alpha=0$ when 
$r\rightarrow 0$. In the opposite limit $r\rightarrow \infty$ we obtain
\begin{equation}
\tanh\alpha_\infty= \frac{1}{s^{-1}+1+s}.
\label{yinfinity}
\end{equation}
On inserting this back into Eq. (\ref{betaTMS}) we find that 
\begin{equation}
\lim_{r\rightarrow\infty}\tanh \beta=1.
\end{equation}
It is easy to check that the equations for $\alpha$ and $\beta$ have 
finite solutions for any finite $r$. 
In the limit $r\rightarrow\infty$, $\alpha$ approaches a
finite asymptotic value, cf. Eq. (\ref{yinfinity}), 
while $\beta$ grows to infinity.

Suppose now that we want to implement the beam splitter 
transformation (\ref{SBS}).  The calculations of the parameters $\alpha$
and $\beta$ parallel those for the two-mode squeezer. From the condition
$S_{12}=-S_{21}$ we express $\beta$ in terms of $\alpha$:
\begin{equation}
\tanh\beta=\frac{-2\sinh\alpha\cosh\alpha}%
{\cosh^2\alpha+(s^{-2}-1+s^2)\sinh^2\alpha}.
\label{AMPbetaBS}
\end{equation}
Since $(s^{-2}-1+s^2)\geq 1$, it follows that 
$|\tanh \beta|\leq |\tanh(2\alpha)|$. From the condition 
$S_{12}/S_{11}=\tan\phi$ we obtain quadratic equation for $\tanh\alpha$
leading to,

\begin{equation}
\tanh\alpha=\frac{(s^{-1}-s)-{\rm
sign}(s^{-1}-s)\sqrt{\frac{s^{-2}-1+s^2}{\cos^2\phi} -1}}%
{(s^{-2}-1+s^2)\tan\phi}.
\label{AMPalphaBS}
\end{equation}
The ${\rm sign}$ function in the above formula selects the root 
that yields the correct limit $\alpha=0$ when $\phi\rightarrow 0$.
The formula (\ref{AMPalphaBS}) is applicable in the interval $\phi\in[0,\pi/2]$. In the
limiting case $\phi=\pi/2$ we have
\begin{equation}
\tanh\alpha_{\pi/2}=\frac{{\rm sign}(s-s^{-1})}{\sqrt{s^{-2}-1+s^2}},
\end{equation}
which implies that $|\tanh\alpha_{\pi/2}|<1$  iff $s\neq 1$. Furthermore, it can
be shown that $\alpha$ is a monotonic function of $\phi$ in the interval
$[0,\pi/2]$. We can thus conclude that 
with the interaction Hamiltonian of the amplifier type (\ref{Hplus}) 
we can implement any beam splitter transformation (\ref{SBS}) with the mixing
angles in the interval $[0,\pi/2]$, which includes the two important cases of a
balanced beam splitter ($\phi=\pi/4$) and the swap ($\phi=\pi/2$).  
It follows
from the expressions (\ref{AMPbetaBS}) and (\ref{AMPalphaBS}) that the
simulation becomes more and more time consuming 
for Hamiltonians close to the two-mode squeezing Hamiltonian
$H_{\rm TMS}=x_Ap_B+p_Ax_B$, i.e., when $s\rightarrow 1$.

\subsection{Beam splitter-like Hamiltonians}

Having derived the gate synthesis parameters for the Hamiltonians (\ref{Hplus}), we
proceed to the interaction Hamiltonians leading to oscillatory dynamics,
\begin{equation}
H_1= x_A p_B - s^2 p_A x_B, \qquad
H_2=-s^2x_A p_B + p_A x_B.
\label{Hminus}
\end{equation}
The $\textbf{S}$ matrices  associated with these Hamiltonians read
\begin{equation}
\bm{S}_{1}^{-}(t)=
\left(
\begin{array}{cc}
\cos(st) & -s\sin(st) \\ 
\frac{1}{s}\sin(st) & \cos(st)
\end{array}
\right),
\label{Sminusone}
\end{equation}
\begin{equation}
\bm{S}_{2}^{-}(t)=
\left(
\begin{array}{cc}
\cos(st) & \frac{1}{s}\sin(st) \\ 
-s\sin(st) & \cos(st)
\end{array}
\right).
\label{Sminus}
\end{equation}

We shall not repeat the details of the derivations of the parameters $\alpha$
and $\beta$ and we only summarize the results here. The beam splitter
transformation can be accomplished by the following choice:
\begin{equation}
\tan\beta=\frac{-2\sin\alpha\cos\alpha}{\cos^2\alpha+(s^{-2}+1+s^2)\sin^2\alpha}
\end{equation}
and
\begin{equation}
\tan\alpha=\frac{s^{-1}+s-\sqrt{(s^{-2}+1+s^2)(\cos\phi)^{-2}+1}}%
{(s^{-2}+1+s^2)\tan\phi}.
\end{equation}
This reveals that simulation of any  beam splitter with $\phi\in[0,\pi/2]$ 
is possible and the parameters satisfy $|\alpha|\leq \pi/2$ and $|\beta|\leq
\pi/2$. 

Consider now the two-mode squeezing operation (\ref{STMS}). After some algebra,
one obtains
\begin{equation}
\tan\beta=\frac{2\sin\alpha\cos\alpha}{\cos^2\alpha-(s^{-2}-1+s^2)\sin^2\alpha}
\end{equation}
and
\begin{equation}
\tan\alpha=\frac{s^{-1}-s-{\rm sign}(s^{-1}-s)
\sqrt{\frac{s^{-2}-1+s^2}{\cosh^2 r}-1}}{(s^{-2}-1+s^2)\tanh r}.
\label{BSalphaTMS}
\end{equation}
The function $\tan \alpha$ must be real which implies that the term under the 
square root must be non-negative. This constraint, in turn, limits the amount of
two-mode squeezing that can be produced via a three-step protocol 
(\ref{threesteps}). It holds that $r\leq r_{\rm th}$ where
\begin{equation}
\cosh r_{\rm th}=\sqrt{s^{-2}-1+s^2}.
\end{equation}
The origin of this bound lies in the fact that the
dynamics governed by the Hamiltonians (\ref{Hminus}) and captured by 
the matrices (\ref{Sminusone}) and (\ref{Sminus}) is oscillatory and 
fully periodic 
with period $2\pi/s$. Squeezing above the threshold $r_{\rm th}$ can 
be achieved only if we concatenate several three-step protocols.
It thus appears that the amplifier-like Hamiltonians (\ref{Hplus})
are in certain sense more versatile than the beam-splitter like Hamiltonians 
(\ref{Hminus}), because the former allow to implement any two-mode squeezing gate
and also any beam splitter with  $\phi\in[0,\pi/2]$ via 
a three-step protocol (\ref{threesteps}).

\section{Conclusions}

In this paper we have addressed the problem of gate synthesis for continuous
variable systems. We have assumed that two single-mode systems $A$ and $B$
interact via a quadratic Hamiltonian. We have studied how to implement a
unitary symplectic gate $G$ with the use of the interaction Hamiltonian $H$ 
as a resource. The gate synthesis protocol consists of a sequence of evolutions
governed by $H$ and followed by fast local phase shifts applied to the systems
$A$ and $B$. We have focused on the gate simulation protocols that involve the
minimal necessary number of steps, because these protocols require a low number
of local control operations which is important from the experimental point of
view. We have shown that a three-step protocol suffices for simulation of the
two-mode squeezer as well as a beam splitter. For the specific case of the
Hamiltonian (\ref{Hxp}) we have also established a four step implementation of a
single-mode squeezer. Our results are applicable to any physical systems coupled
via quadratic Hamiltonians. In particular, the gate synthesis protocols 
proposed in the present paper may find applications in the 
experiments where light interacts  with atomic ensembles via a Kerr-like
coupling \cite{Kuzmich00,Julsgaard01,Schori02}.

\acknowledgments

I would like to thank S. Massar, E.S. Polzik, N.J. Cerf and F. Grosshans 
for helpful discussions.
I acknowledge financial support from the Communaut\'e Fran\c{c}aise de
Belgique under grant ARC 00/05-251, from the IUAP programme of the Belgian
governement under grant V-18, from the EU under projects RESQ
(IST-2001-35759) and CHIC (IST-2001-32150)  and from the 
grant LN00A015  of the Czech Ministry of Education.


\begin{thebibliography}{99}

% Hamiltonian simulation

\bibitem{Dur01}
W. D\"{u}r, G. Vidal, J.I. Cirac, N. Linden, and S. Popescu,
Phys. Rev. Lett. {\bf 87}, 137901 (2001).


\bibitem{Bennett02}
C.H. Bennett, J.I. Cirac, M.S. Leifer, D.W. Leung, N. Linden, S. Popescu, 
and G. Vidal, Phys. Rev. A \textbf{66}, 012305 (2002).

\bibitem{Dodd02}
J.L. Dodd, M.A. Nielsen, M.J. Bremner, and R.T. Thew,
Phys. Rev. A \textbf{65}, 040301 (2002).

\bibitem{Nielsen02}
M.A. Nielsen, M.J. Bremner, J.L.Dodd, A.M. Childs, and C.M. Dawson, 
Phys. Rev. A \textbf{66}, 022317 (2002).

\bibitem{Wocjan02}
P. Wocjan, M. R\"{o}tteler, D. Janzing, and T. Beth,
Phys. Rev. A \textbf{65}, 042309 (2002).

\bibitem{Vidal02}
G. Vidal and J.I. Cirac,
Phys. Rev. A \textbf{66} 022315 (2002).

% Gates simulation

\bibitem{Khaneja01}
N. Khaneja, R. Brockett, and S.J. Glaser, 
Phys. Rev. A \textbf{63}, 032308 (2001).

\bibitem{Vidal02b}
G. Vidal, K. Hammerer, and J.I. Cirac, 
Phys. Rev. Lett. \textbf{88}, 237902 (2002).

\bibitem{Hammerer02}
K. Hammerer, G. Vidal, and J.I. Cirac, quant-ph/0205100.

\bibitem{Masanes02}
L. Masanes, G. Vidal, and J.I. Latorre, 
quant-ph/0202042.

\bibitem{Bremner02}
M.J. Bremner, C.M. Dawson, J.L. Dodd, A. Gilchrist, A.W. Harrow, 
D. Mortimer, M.A. Nielsen, and T.J. Osborne,
Phys. Rev. Lett. \textbf{89}, 247902 (2002). 


\bibitem{Zhang02}
J. Zhang, J. Vala, S. Sastry, and K.B. Whaley, quant-ph/0212109.


% CV Hamiltonian simulation 
\bibitem{Kraus02}
B. Kraus, K. Hammerer, G. Giedke, and J.I. Cirac,
quant-ph/0210136.


% Polzik and company atomic cloud experiments

\bibitem{Kuzmich98}
A. Kuzmich, N.P. Bigelow, and L. Mandel, 
Europhys. Lett. \textbf{42}, 481 (1998).


%squeezing
\bibitem{Kuzmich00}
A. Kuzmich, L. Mandel, and N.P. Bigelow,
Phys. Rev. Lett. \textbf{85}, 1594 (2000).

\bibitem{Kuzmich00theory}
A. Kuzmich and E.S. Polzik, 
Phys. Rev. Lett. \textbf{85}, 5639 (2000).

\bibitem{Duan00}
L.M. Duan, J.I. Cirac, P. Zoller, and E.S. Polzik, 
Phys. Rev. Lett. \textbf{85}, 5643 (2000).

% entanglement
\bibitem{Julsgaard01}
B. Julsgaard, A. Kozhekin, E.S. Polzik,
Nature (London) \textbf{413}, 400 (2001).

% transfer of light onto atomic cloud
\bibitem{Schori02}
C. Schori, B. Julsgaard, J.L. Sorensen, and E.S. Polzik,
Phys. Rev. Lett. \textbf{89}, 057903 (2002).

\bibitem{Kuzmichbook}
A. Kuzmich and E.S. Polzik,
in {\em Quantum Information with Continuous Variables},
edited by S.L. Braunstein and A.K. Pati 
(Kluwer Academic, to be published). 

\bibitem{Polzikprivate}
E.S. Polzik, private communication.


% symplectic group
\bibitem{Simon94}
R. Simon, N. Mukunda, and B. Dutta,
Phys. Rev. A \textbf{49}, 1567 (1994).

\bibitem{Arvind95}
Arvind, B. Dutta, N. Mukunda, and R. Simon,
Phys. Rev. A 52, \textbf{1609} (1995).



\bibitem{Braunstein99}
S. L. Braunstein, quant-ph/9904002.


% sbscheme for couplers

\bibitem{Fiurasek00}
J. Fiur\'{a}\v{s}ek and J. Pe\v{r}ina,
Phys. Rev. A \textbf{62}, 033808 (2000).

\bibitem{Rehacek02}
J. \v{R}eh\'{a}\v{c}ek, L. Mi\v{s}ta, Jr., J. Fiur\'{a}\v{s}ek, 
and J. Pe\v{r}ina, 
Phys. Rev. A \textbf{65}, 043815 (2002). 



\end{thebibliography}
\end{document}